\documentclass[10 pt]{article}
\usepackage{epsfig}
\begin{document}

\title{\bf Reply to `On a recent proposal of faster than light quantum communication'}

\author{A.Y. Shiekh\footnote{\rm shiekh@dinecollege.edu} \\
             {\it Din\'{e} College, Tsaile, Arizona, U.S.A.}}

\date{}

\maketitle

\abstract{In a recent paper \cite{Shiekh FTL} the author proposed the possibility of an experiment to perform faster-than-light communication via the collapse of the quantum wave-function. This was analyzed by Bassi and Ghirardi \cite{Bassi and Ghirardi}, and it is believed that this analysis itself merits a detailed examination.}

\baselineskip .5 cm

\section{The original proposal}
The proposed device is based upon the conservation of a particle in quantum theory (unitarity), and is founded upon the following apparatus \cite{Shiekh FTL}

\begin{center}
\includegraphics[scale=.35]{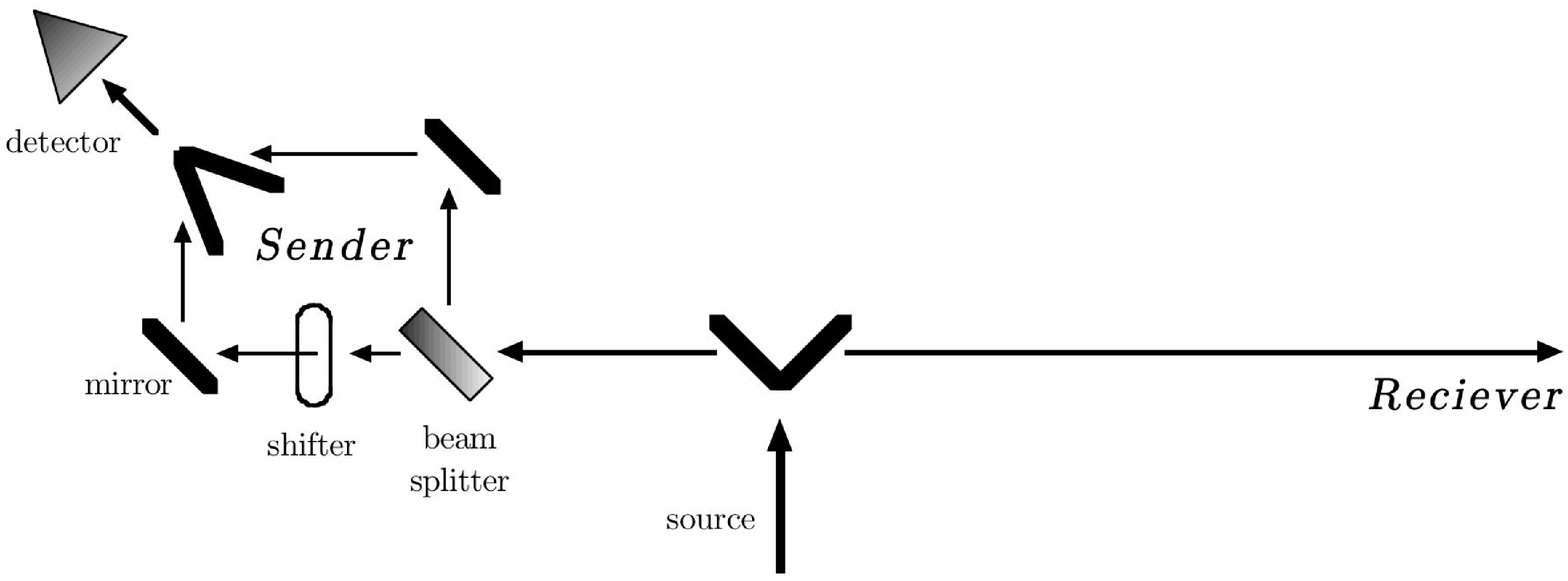}
\end{center}
\begin{center}
Quantum Transmitter
\end{center}
where a beam splitting mechanism breaks a single particle wave-function into two arms that can be widely separated, and then again splits and recombines one of the two resulting arms.

The recombination can be arranged to constructively, or destructively interfere, depending on a phase shifter in one of the two sub-paths.

If the sender chooses to arrange for constructive interference then some of the particles will be `taken up' by the sender, but less if destructive interference is arranged, and in this way the intensity of the receiver's beam might be controlled. So a faster-than-light transmitter of information (but not energy or matter) might be possible.

\section{The counter analysis}
The proposal outlined above was recently analyzed by Bassi and Ghirardi \cite{Bassi and Ghirardi}, and while they seem to agree that the arranged for destructive interference will induce a re-unitarization, they seem convinced that this will only happen for the one arm where interference is being arranged, while the original proposal re-normalized the entire state.

However, a `localized' re-normalization, rather than avoiding a faster-than-light communicator can itself not only result in one, but has some other possibly unexpected consequences, as detailed below.

\subsection{A faster-than-light communicator, when there should be none}
All authors agree that the wave function after the second splitting, but before recombination, is described by
\begin{equation}
\frac{\left| \phi_{s_1} \right>}{\sqrt{2}} + \frac{\left| \phi_{s_2} \right>}{\sqrt{2}} + \left| \phi_{r} \right>
\end{equation}
where $s_1$ and $s_2$ are the two sender's arms and $r$ is the receiver; overall re-normalization of each state is understood, but left out for clarity.

Now suppose that a measurement is made first in one of the sender's sub-arms, and then the other. In the case where the particle was {\it not} found in this first measurement one would get, for the Bassi Ghirardi approach of re-normalizing the sender's arm alone
\begin{equation}
\left| \phi_{s_2} \right> + \left| \phi_{r} \right>
\end{equation}
as opposed to the original re-normalizing of the entire state, where one would get
\begin{equation}
\frac{\left| \phi_{s_2} \right>}{\sqrt{2}} + \left| \phi_{r} \right>
\end{equation}

These two approaches naturally lead to differing predictions for the probability of finding the particle in the sender's arm after both measurements are made. For the first (component re-normalization) one would get a predicted probability of 1/4 of finding the particle in the first sub-arm and then 1/2 of finding it in the second sub-arm (if it was not found in the first), yielding a total probability of 1/4 + 1/2 (1 - 1/4), an unexpected 62.5\% result. On the other hand, for the overall re-normalization, the same calculation 1/4 + 1/3 (1 - 1/4) yields the expected result of 50\%.

So the first approach (however justified) would immediately give rise to a faster-than-light communicator, as the sender could then either opt to measure the main arm (with a resulting 50\% chance of seeing the particle), or instead perform measurements of the two sub-arms for a predicted 62.5\% chance of seeing the particle, so reducing the receiver's intensity if this is done for a beam of particles.

While, for the re-normalization of the entire state, as done for the original proposal, there is no possibility of a faster-than-light communicator in this case, which seems more reasonable.

\subsection{Component re-normalization}
Further problems result from this component re-normalization approach, in that if one uses it to analyze the case of interference (in the previous analysis we elected to measure before interference took place), then according to Bassi and Ghirardi's section III an area of destructive interference where the particle is less likely to be found does indeed occur, but is accompanied by a boosting of the wave-function just outside of this area to maintain unitarity (FIG 2b). This in itself would constitute a faster-than-light communicator, albeit over a rather limited range. In the original proposal it is the entire state, not just part, that is boosted by the conservation of unitarity (which is not violated in either work).

On a side note, perfect cancellation is certainly not possible in practice, and only needs to be partial in this application.

\subsection{General Proof}
While there is reference to proofs of `full generality' against faster-than-light communication, it is at the same time conceded that such proofs are for two particle entangled states and that the present case does not fall under such `general' proofs.

\subsection{Super-luminal communication}
Section VI (the Conclusions) claims that such a faster-than-light communication device would allow for the synchronizing of clocks and so imply the existence of absolute time. However, in general, since the sender and receiver would have a relative velocity, their respective clocks would be running at differing rates, and so could not run in synchrony.

\section{Conclusion}
While the possibility of constructing a faster-than-light communicator is highly unlikely, it is not believed that the work of Bassi and Ghirardi has yet located the flaw in the present proposal. As demonstrated here, their approach of re-normalizing part (and not all) of the state itself gives rise to a faster-than-light communicator, and so cannot constitute a proof against such a proposal.

The mechanism of `magnification' through re-unitarization (seen in both works) can also be used to pick out desirable components in the case of quantum computation \cite{Shiekh CC1, Shiekh CC2}, although in that instance the cancellation needs to be near perfect, and achievable in polynomial time (as can be easily demonstrated).

\section{Acknowledgements}
I would like to thank Professor Giancarlo Ghirardi for extensive discussions on the above issues.


\begin{thebibliography}{99}
\bibitem{Shiekh FTL} A. Y. Shiekh, {\it Faster than Light Quantum Communication}, arXiv:0710.1367v1
\bibitem{Bassi and Ghirardi} A. Bassi and G. C. Ghirardi, {\it On a recent proposal of faster than light quantum communication}, arXiv:0711.4538v1
\bibitem{Shiekh CC1} A. Y. Shiekh, Int. Jour. Theo. Phys., 45, 2006, 1646 [arXiv:cs.CC/0507003]
\bibitem{Shiekh CC2} A. Y. Shiekh, arXiv:quant-ph/0611052, arXiv:0704.2033
\end{thebibliography}
\end{document}